\documentclass[preprintnumbers, prd, twocolumn, showpacs, floatfix,
preprintnumbers, letterpaper, superscriptaddress,nofootinbib]{revtex4}
\usepackage{amsfonts}
\usepackage{eurosym}
\usepackage[dvips]{graphicx}
\usepackage{epsf}
\usepackage{amsmath}
\usepackage{amssymb}
\usepackage{graphicx}
\usepackage{dcolumn}
\usepackage{bm}
\usepackage{color,wasysym}
\usepackage{amsmath,amsfonts,amssymb}
\usepackage{latexsym}
\usepackage{dcolumn}
\usepackage{graphicx,epsfig}
\usepackage{amsthm}
\usepackage{color}
\usepackage{hyperref}
\begin{document}

\title{{\bf Black Hole Solutions in Rastall Theory}}
\author{Y. Heydarzade}
\email{heydarzade@azaruniv.edu; Corresponding author}
\affiliation{Department of Physics, Azarbaijan Shahid
Madani University, Tabriz,
53714-161,
Iran}
\affiliation{ Research Institute for Astronomy and Astrophysics of Maragha (RIAAM),
Maragha
55134-441, Iran}
\author{ H. Moradpour}
\email{h.moradpour@riaam.ac.ir}
\affiliation{ Research Institute for Astronomy and Astrophysics of Maragha (RIAAM),
Maragha
55134-441, Iran}
\author{F. Darabi}
\email{f.darabi@azaruniv.edu}
\affiliation{Department of Physics, Azarbaijan Shahid
Madani University, Tabriz, 53714-161, Iran}
\affiliation{ Research Institute for Astronomy and Astrophysics of Maragha (RIAAM), Maragha 55134-441, Iran}
\begin{abstract}
The Reissner-Nordstr\"om black hole  solution in a generic cosmological
constant background   in the the context of Rastall gravity is obtained.
It is shown that the cosmological constant arises naturally from the consistency
of the non-vacuum field equations of the Rastall theory for a spherical symmetric spacetime, rather than its {\it
ad-hoc} introduction in the usual Einstein and Einstein-Maxwell field equations.
The  usual Reissner-Nordstr\"om,  Schwarzschild and
Schwarzschild-(anti)de Sitter black hole solutions in the framework of this
theory are also addressed as the special independent subclasses of the obtained general
solution.\\
\\
Keywords:{ Reissner-Nordstr\"om black hole,  Rastall theory of gravity.}
\end{abstract}
\pacs{04.50.-h, 04.20.Jb, 04.70.-s}

\maketitle


\section{Introduction}
In the curvature-matter coupling theory of gravity
\cite{od1,od2,cmc,cmc1,cmc2}, the geometry and matter fields  are
coupled to each other in a non-minimal way meaning that the
Lagrangian of the system is not a simple summation of the Lagrangian
of geometry and the matter fields. Due to this non-minimal coupling,
the geometry and  matter fields are affected by their mutual
changes, so the ordinary energy-momentum conservation law may be
invalid in this context. In fact, in this theory, the divergence of
the energy-momentum tensor is related to the changes of the Ricci
scalar \cite{od1,od2,cmc,cmc1,cmc2}.

The idea of coupling of the matter and geometry fields in a
non-minimal way comes back to P. Rastall \cite{rastall}, who
challenged the energy-momentum conservation law in the curved
spacetime for the first time, a hypothesis supported by the particle
creation during the cosmos evolution
\cite{motiv1,motiv2,motiv3,motiv4, rw1}. In this way, it was shown
by Smalley that a prototype of Rastall''s theory of gravity, in
which the divergence of the energy-momentum tensor is proportional
to the gradient of the Ricci scalar, can be derivable from a
variational principle \cite{smal}. It was discussed that both the
proportionality factor and the unrenormalized gravitational constant
are  covariantly constant, but not necessarily constant. In this
regard, this theory appears as a gravitational theory with variable
gravitational constant.  In a cosmological context, this theory is
in agreement with some observational data and leads to some
interesting results \cite{al1,al2}. As an instance, the evolution of
small dark matter fluctuations is identical with that of the
$\Lambda CDM$ model, but in this' theory dark energy is able to be
clustered. This feature result in an evolution of dark matter
inhomogeneities in a nonlinear regime which is different from the
one in the standard $CDM$ model \cite{prd}. Thermodynamical aspects
of this model in a flat FLRW universe are studied in
\cite{plb,msal}. This theory is also studied in the context of the
G\"{o}del-type universe \cite{san}.

Moreover, some static spherically symmetric solutions of the Rastall
field equations have  been obtained. For example, by special setting
of the parameters of the Rastall theory,  a vacuum solution
possessing the same structure as the Schwarzschild-de Sitter
solution in the general relativity theory obtained with a
cosmological constant playing the role of source \cite{oliveira}. In
another work,  the problem of finding static spherically symmetric
spaces in Rastall''s theory in the presence of a free or self
interacting scalar field is studied \cite{r0}. It is found that some
exact solutions can be obtained in which  some of these solutions
are the same as the solutions which are obtained in the context of
k-essence theory. Also, the asymptotically flat traversable wormhole
solutions are investigated in \cite{hms} where it is shown that the
parameters of Rastall theory affect the parameters of the wormhole
spacetime.

In this work, our aim  is to find general static spherical symmetric
black hole solutions in the context of the Rastall theory of gravity
as a special class of the extended theories of gravity \cite{BHs}.
The paper is organized as follows. In the section II, we have a
review on the Rastall theory of gravity and its general remarks. In
section III, we investigate the spherical symmetric black hole
solutions in the Rastall theory  equipped by the electromagnetic
sector. Then, we obtain the charged black hole  in the cosmological
constant background. Finally, in section IV, we represent our
concluding remarks.

\section{Rastall Theory of Gravity}
Based on the Rastall's hypothesis \cite{rastall}, for a spacetime
with Ricci scalar $R$ filled by a source of $T^{\mu \nu}$ we have
\begin{eqnarray}\label{rastal}
{T^{\mu \nu}}_{;\mu}=\lambda R^{,\nu},
\end{eqnarray}
where $\lambda$ is the Rastall parameter. The Rastall field  equation can be
written as
\begin{eqnarray}\label{r1}
G_{\mu \nu}+\gamma g_{\mu \nu}R=\kappa T_{\mu \nu},
\end{eqnarray}
where $\gamma=\kappa\lambda$ and $\kappa$ is the Rastall
gravitational coupling constant. Equivalently, one can rewrite this
equation in the following form
\begin{equation}\label{ein}
G_{\mu \nu}=\kappa S_{\mu\nu},
\end{equation}
where
\begin{equation}\label{senergy}
S_{\mu\nu}=T_{\mu\nu}-\frac{\gamma T}{4\gamma-1}g_{\mu\nu},
\end{equation}
is the effective energy-momentum tensor in which
\begin{eqnarray}\label{scomp}
&&{S^{0}}_{0}\equiv-\rho^{e}=-\frac{(3\gamma-1)\rho+\gamma(p_r+2p_t)}{4\gamma-1},\nonumber\\
&&{S^{1}}_{1}\equiv
p^{e}_r=\frac{(3\gamma-1)p_r+\gamma(\rho-2p_t)}{4\gamma-1},\nonumber\\
&&{S^{2}}_{2}={S^{3}}_{3}\equiv
p^{e}_t=\frac{(2\gamma-1)p_t+\gamma(\rho-p_r)}{4\gamma-1}.
\end{eqnarray}
Here, $\rho^e$, $p^e_r$ and $p^e_t$ are the effective density and
pressures corresponding to the source with the energy-momentum
tensor ${T^{\mu}}_{\nu}$, respectively \cite{hms}. The usual
Einstein general relativity can be recovered in the appropriate
limit of $\lambda\rightarrow0$. Also, it is apparent that for a
traceless energy-momentum source, such as the electromagnetic
source, we have $S_{\mu\nu}=T_{\mu\nu}$ which leads to $G_{\mu
\nu}=\kappa T_{\mu\nu}$. This means that the Einstein solutions for
$T=0$, or equivalently $R=0$, are also valid in the Rastall theory
of $\kappa$ \cite{al1,r0}. Moreover, it should be mentioned that both of the parameter
cases $\gamma=\frac{1}{4}$ and $\gamma=\frac{1}{6}$
are not allowed as discussed in \cite{msal}.  Using the Newtonian limit, it is shown
that the Rastall parameter $\lambda$ and Rastall gravitational
coupling constant $\kappa$ diverge for $\gamma=\frac{1}{4}$ and
$\gamma=\frac{1}{6}$, respectively, which does not make a physical
sense \cite{msal}.

In the following, we will work in the units where $\kappa=1$,
leading to $\gamma=\lambda$, and therefore we have the following
Rastall field equation
\begin{equation}\label{ein22}
G_{\mu \nu}=S_{\mu\nu},
\end{equation}
where
\begin{equation}\label{senergy22}
S_{\mu\nu}=T_{\mu\nu}-\frac{\lambda
T}{4\lambda-1}g_{\mu\nu}.
\end{equation}

\section{ Black Hole Solutions in Rastall-Maxwell Theory}
In this section, we construct Rastall-Maxwell theory in order to find charged black hole solutions.  We consider the following
static spherical symmetric
metric in the context of Rastall theory
possessing the  modified field equation (\ref{ein22})
\begin{equation}\label{metric}
ds^2 =-e^{\mu(r)}dt^2+e^{\nu(r)}dr^2+r^2 d\Omega_{2}^2 ,
\end{equation}
with the total energy-momentum tensor given by
\begin{equation}\label{t}
T_{\mu\nu}=T^{*}_{\mu\nu}+E_{\mu\nu},
\end{equation}
where
\begin{equation}\label{T*}
T^{*}_{\mu\nu}=(\rho+p)u_{\mu}u_{\nu}+pg_{\mu\nu},
\end{equation}
is the non-vanishing trace part of the total energy momentum tensor possessing the barotropic equation of state $p=\omega\rho$, and $E_{\mu\nu}$ is the trace-free Maxwell tensor given by
\begin{equation}\label{E}
E_{\mu\nu}=2\textcolor[rgb]{1,0,0.501961}{}\left(F_{\mu\alpha}{F_{\nu}}^{\alpha}-
\frac{1}{4}g_{\mu\nu}F^{\alpha\beta}F_{\alpha\beta}\right),
\end{equation}
where $F_{\mu\nu}$ is the antisymmetric Faraday tensor satisfying
the vacuum Maxwell equations
\begin{eqnarray}\label{max}
&&{F^{\mu\nu}}_{;\mu}=0,\nonumber\\
&&\partial_{[\sigma} F_{\mu\nu]}=0.
\end{eqnarray}
Using the equations (\ref{senergy22}) and (\ref{t}), we obtain the effective energy momentum tensor $S_{\mu\nu}$
as
\begin{equation}\label{senergy}
S_{\mu\nu}=T^{*}_{\mu\nu}-\frac{\lambda
T^*}{4\lambda-1}g_{\mu\nu} +E_{\mu\nu}.
\end{equation}
On the other hand, spherical symmetry implies that the only non-zero
components of $F^{\mu\nu}$ are $F^{01}=- —F^{10}$. Then, from the equations in (\ref{max}),
one obtains
\begin{equation}\label{f}
F^{01}=\frac{Q}{r^2}.
\end{equation}
Using the equations (\ref{metric}), (\ref{E}) and (\ref{f}), the only non-vanishing components of ${E^{\mu}}_{\nu}$ will be
\begin{equation}
{E^{\mu}}_{\nu}=\frac{Q^2}{ r^4}~e^{\mu+\nu}~diag(-1,-1,1,1).
\end{equation}
Then, the effective energy momentum  tensor components  take the following
form\begin{eqnarray}\label{scomp1}
&&{S^{0}}_{0}=-\frac{(3\lambda-1)\rho+3\lambda p}{4\lambda-1}-\frac{Q^2}{  r^4}e^{\mu+\nu},\nonumber\\
&&{S^{1}}_{1}=\frac{(\lambda-1)p+\lambda\rho}{4\lambda-1}-\frac{Q^2}{  r^4}e^{\mu+\nu},\nonumber\\
&&{S^{2}}_{2}={S^{3}}_{3}=\frac{(\lambda-1)p+\lambda\rho}{4\lambda-1}+\frac{Q^2}{  r^4}e^{\mu+\nu}.
\end{eqnarray}
Then, the $00$ component of the Rastall field equation (\ref{ein}) yields
\begin{equation}\label{00}
\frac{e^{-\nu}}{r^2}(1-r\nu^{\prime}-e^\nu)=-\frac{(3\lambda-1)\rho+3\lambda p}{4\lambda-1}-\frac{Q^2}{ r^4}e^{\mu+\nu},
\end{equation}
and $11$ component gives
\begin{equation}\label{11}
\frac{e^{-\nu}}{r^2}(1+\mu^{\prime}r-e^\nu)=\frac{(\lambda-1)p+\lambda\rho}{4\lambda-1}-\frac{Q^2}{ r^4}e^{\mu+\nu}.
\end{equation}
Finally, the $22$ and $33$ components lead to
\begin{eqnarray}\label{22}
&&\frac{e^{-\nu}}{4r}(2\mu^{\prime}-2\nu^{\prime}+r\mu^{\prime}\nu^{\prime}+2\mu^{\prime\prime
}r+{\mu^{\prime}}^2 r)\nonumber\\
&&~~~~~~~~~~~~~~=\frac{(\lambda-1)p+\lambda\rho}{4\lambda-1}+\frac{Q^2}{ r^4}e^{\mu+\nu}.
\end{eqnarray}
One can consider the usual special class of the spherical symmetric metrics as
$\nu(r)=-\mu(r)$.
Consequently, by setting $\nu(r)=-ln(1-f(r)),$  the $00$ component of the Rastall field equation (\ref{00}) yields
\begin{equation}\label{00*}
-f(r)-rf^{\prime}(r)=-\frac{(3\lambda-1)\rho+3\lambda p}{4\lambda-1}r^2-\frac{Q^2}{ r^2},
\end{equation}
and $11$ component gives
\begin{equation}\label{11*}
-f(r)-rf^{\prime}(r)=\frac{(\lambda-1)p+\lambda\rho}{4\lambda-1}r^{2}-\frac{Q^2}{ r^2}.
\end{equation}
Finally, the $22$ and $33$ components lead to
\begin{equation}\label{22*}
-rf^{\prime}(r)-\frac{1}{2}r^2f^{\prime\prime}=\frac{(\lambda-1)p+\lambda\rho}{4\lambda-1}r^{2}+\frac{Q^2}{ r^2}.
\end{equation}
It is clear that the left hand sides of the equations (\ref{00*}) and (\ref{11*})
are the same. Then, the consistency for the right hand sides of
these equations  requires that  $4 \lambda=1$ or the matter
field described by (\ref{T*}) with $\rho\neq 0$ takes the equation of
state parameter $\omega=-1$. For the case of $\rho=p=0$, these equations
are trivially consistent. The first possibility,
i.e $4\lambda=1, $ is ruled out based on
the discussions in \cite{hms}. Regarding the second possibility,
the black hole surrounding matter has an effective behavior as the same as the cosmological
constant.  It is easy to verify that the equations (\ref{00*}) and (\ref{22*})
are satisfied by the  metric function $f(r)$ given by
\begin{equation}
f(r)=\frac{2M}{r}-\frac{Q^2}{r^2}+h(r),
\end{equation}
where $M$ and $Q$ are the black hole mass and charge respectively, and  $h(r)$ is a function which should satisfy the two following equations  resulting from
the equations (\ref{11*}) and (\ref{22*})
\begin{equation}\label{g}
-h(r)-rh^{\prime}(r)=\frac{\rho(r)}{4\lambda-1}r^{2},
\end{equation}
and
\begin{equation}\label{g1}
-rh^{\prime}(r)-\frac{1}{2}r^{2}h^{\prime\prime}(r)=\frac{\rho(r)}{4\lambda-1}r^{2}.
\end{equation}
 One can consider a general situation where the energy density $\rho(r)$  consists of two parts,
one constant part and another asymptotically vanishing part, i.e $\rho(r)=\rho_{*}+\frac{\sigma_*}{r^n}$
where $\rho_*$ and $\sigma_*$ are positive constants with the dimension of
energy density and energy density times $(Length)^n$, respectively.
Then,  regarding the equation (\ref{g}), the solution  for $h(r)$
function will be
\begin{equation}
h(r)=\alpha r^2 + \frac{\sigma}{r^{m}},
\end{equation}
where $\sigma=\frac{\sigma_*}{(m-1)(4\lambda-1)}$ and $\alpha=-\frac{1}{3}\frac{1}{4 \lambda-1}\rho_*$ are integration constants.  Then, substituting $h(r)$
in  (\ref{g1}), one finds two possibilities for $m,$ for the consistency of
the equation (\ref{g1}), as $m=1$ and $m=-2$.
Because $\sigma$ diverges for the case of $m=1$, this case has no physical
meaning and can be discarded. For the case of $m=-2$ or equivalently $n=0$, one arrives at $h(r)=\Lambda
r^2$ where $\Lambda$ is the cosmological constant as $\Lambda=-\frac{1}{3}\frac{1}{4 \lambda-1}(\rho_*+\sigma_{*})\equiv -\frac{1}{3}\frac{1}{4 \lambda-1}\rho_0$.
Then, the only possibility is  the constant energy density $\rho=\rho_0$ representing the vacuum energy or the cosmological constant
which we previously obtained its effective equation of state parameter $\omega=-1$ using the consistency
of the field equations (\ref{00*})-(\ref{22*}) of the Rastall theory of gravity. Also, note that here the $\lambda=1/4$ parameter leads to a divergent cosmological constant and then is not physically acceptable as we previously mentioned. Finally, the metric will take the following form
\begin{eqnarray}\label{metric11}
ds^2 &=&-\left(1-\frac{2M}{r}+\frac{Q^2}{r^2} -\Lambda r^2  \right)dt^2\nonumber\\ &&+\frac{dr^2}{1-\frac{2M}{r}+\frac{Q^2}{r^2} -\Lambda r^2}+r^2 d\Omega_{2}^2.
\end{eqnarray}
Then, we summarize our results in the following points:
\begin{itemize}
\item The  metric (\ref{metric11}) represents the Reissner-Nordstr\"om black hole  in the cosmological
constant background  which we obtained in the the context of the Rastall gravity.
 \item Note that the cosmological constant arises from the consistency
of the non-vacuum field equations of the Rastall theory for a spherical symmetric
spacetime rather than its {\it
ad-hoc} introduction to usual Einstein and Einstein-Maxwell field equations. It is also interesting that
in the framework of this theory, the Rastall coupling $\lambda$ affects the value of
the cosmological constant. Regarding the weak energy condition, i.e $\rho_0\geq0$,
for $\lambda<\frac{1}{4}$ we have positive cosmological constant
representing de Sitter space while for the $\lambda> \frac{1}{4}$ we have
a negative cosmological constant denoting anti- de Sitter space.
\item Regarding the equations (\ref{00*})-(\ref{22*}), it is easy to check that for  $T^{*}_{\mu\nu}=0$ or equivalently
$\rho=p=0$, we arrive at the usual  Reissner-Nordstr\"om black hole solution
\begin{equation}\label{metric2}
ds^2 =-\left(1-\frac{2M}{r}+\frac{Q^2}{r^2}\right)dt^2+\frac{dr^2}{1-\frac{2M}{r}+\frac{Q^2}{r^2}}+r^2 d\Omega_{2}^2.
\end{equation}

\item Regarding the equations (\ref{00*})-(\ref{22*}), for the  vacuum solution, i.e $T^{*}_{\mu\nu}=E_{\mu\nu}=0$
or equivalently $\rho=p=Q=0$,
we arrive at the Schwarzschild solution
\begin{equation}\label{metric3}
ds^2 =-\left(1-\frac{2M}{r}\right)dt^2+\frac{dr^2}{1-\frac{2M}{r}}+r^2 d\Omega_{2}^2.
\end{equation}
\item Regarding the equations (\ref{00*})-(\ref{22*}), for the the
case of $T^{*}_{\mu\nu}\neq 0$ and  $E_{\mu\nu}=0$ or equivalently $\rho, p\neq0$ and $Q=0$, we recover the Schwarzschild-(anti)de Sitter
solution\begin{equation}\label{metric1}
ds^2 =-\left(1-\frac{2M}{r}-\Lambda r^2  \right)dt^2+\frac{dr^2}{1-\frac{2M}{r}-\Lambda r^2}+r^2 d\Omega_{2}^2.
\end{equation}
Such a solution is obtained as the nontrivial static spherically symmetric vacuum solution of the Rastall theory by solving the equation $R_{\mu\nu}=\frac{1}{4}Rg_{\mu\nu}$
which is resulted by the appropriate setting of the parameters of Rastall
theory  \cite{oliveira}.

\end{itemize}

%
\section{Conclusion}
We obtained the general spherical symmetric Reissner-Nordstr\"om black hole  in the cosmological
constant background   in the the context of the Rastall gravity.  In this
context, the cosmological constant arises naturally from the consistency
of the non-vacuum field equations of the Rastall theory for a spherical symmetric
spacetime rather than its {\it
ad-hoc} introduction in the usual Einstein and Einstein-Maxwell field equations. It is also interesting that
in the framework of this theory, the Rastall coupling $\lambda$ affects the value of
the cosmological constant. For $\lambda<\frac{1}{4}$ we have positive cosmological constant
representing de Sitter space while for the $\lambda> \frac{1}{4}$ we have
a negative cosmological constant denoting anti- de Sitter space.
We also discussed about the usual Reissner-Nordstr\"om,  Schwarzschild and
Schwarzschild-(anti)de Sitter black hole solutions as the independent subclasses
of  the obtained general solution.

\section{Acknowledgments}
This work has been supported financially by Research Institute
for Astronomy and Astrophysics of Maragha (RIAAM) under research project No.
1/4165-84.


\begin{thebibliography}{99}
\bibitem{od1} S. Nojiri, S. D. Odintsov, Phys. Lett. B {\bf599}, 137 (2004).
\bibitem{od2} G. Allemandi, A. Borowiec, M. Francaviglia, S. D. Odintsov, Phys. Rev. D {\bf72}, 063505 (2005).
\bibitem{cmc} T. Koivisto, Class. Quant. Grav. {\bf23}, 4289 (2006).
\bibitem{cmc1} O. Bertolami, C. G. Boehmer, T. Harko, F. S. N. Lobo, Phys. Rev. D {\bf75}, 104016 (2007).
\bibitem{cmc2} T. Harko, F. S. N. Lobo, Galaxies, {\bf2}, 410 (2014).
\bibitem{rastall} P. Rastall, Phys. Rev. D {\bf 6}, 3357 (1972).
\bibitem{motiv1} G. W. Gibbons, S. W. Hawking, Phys. Rev. D {\bf15}, 2738 (1977).
\bibitem{motiv2} L. Parker, Phys. Rev. D {\bf3}, 346 (1971); D {\bf3}, 2546 (1971).
\bibitem{motiv3} L. H. Ford, Phys. Rev. D {\bf35}, 2955 (1987).
\bibitem{motiv4}  N. D. Birrell, P. C. W. Davies, \textit{Quantum Fields in Curved Space} (Cambridge University Press, Cambridge, 1982).
\bibitem{rw1} Ph. Brax, C. van de Bruck, A. Davis, arXiv:0706.1024 (2007).
\bibitem{smal} L. L. Smalley, Il Nuovo Cimento B, {\bf 80}, 1, 42 (1984).
\bibitem{al1} Al-Rawaf, A. S., and Taha, M. O, Phys. Lett. B 366, 69 (1996).
\bibitem{al2} Al-Rawaf, A. S., and Taha, M. O, Gen. Rel. Grav. 28, 935 (1996).
\bibitem{prd} C. E. M. Batista, M. H. Daouda, J. C. Fabris, O. F. Piattella, D. C. Rodrigues, Phys. Rev. D {\bf85}, 084008 (2012).
\bibitem{plb} H. Moradpour, Phys. Lett. B {\bf757}, 187 (2016).
\bibitem{msal} H. Moradpour, I. G. Salako, AHEP, 2016, 3492796 (2016).
\bibitem{san} A. F. Santos, S. C. Ulhoa, Mod. Phys. Lett. A {\bf30}, 1550039 (2015).
\bibitem{oliveira} A. M. Oliveira, H. E. S. Velten, J. C. Fabris, L. Casarini, Phys. Rev. D {\bf93}, 124020 (2016).
\bibitem{r0} K. A. Bronnikov, J. C. Fabris, O. F. Piattella and E. C. Santos, arXiv:1606.06242.
\bibitem{hms} H. Moradpour, N. Sadeghnezhad, arXiv:1606.00846.
\bibitem{BHs}S. Fernando and D. Krug, General Relativity and Gravitation 35, 129 (2003); S. Capozziello,  P. A. Gonzalez, E. N. Saridakis and  Yerko Vasquez, JHEP 1302, 039 (2013); K.C. K. Chan, J. H. Horne and R. B. Mann Phys. lett. B 447, 2–3 (1995); E. Babichev, A. Fabbri, JHEP 07,016 (2014);
A. Sheykhi, Phys. Rev. D 86, 024013 (2012); D. Garfinkle, G. T. Horowitz, and A. Strominger Phys. Rev. D 43, 3140 (1991).
\end{thebibliography}
\end{document}